\documentclass[conference]{IEEEtran}
\IEEEoverridecommandlockouts

\usepackage{amsfonts,amsmath,amssymb,graphicx}
\usepackage{array}
\usepackage{booktabs}
\usepackage{cite}
\usepackage{dsfont}
\usepackage{hyperref}
\usepackage{mwe}
\usepackage[caption=false]{subfig}
\usepackage{textcomp}
\usepackage{xcolor}
\hypersetup{
    colorlinks=true,
    linkcolor=blue,
    filecolor=magenta,      
    urlcolor=cyan,
    pdfpagemode=FullScreen,
    }

\title{Towards Automatic Assessment of Self-Supervised\\ Speech Models using Rank}
\author{\IEEEauthorblockN{Zakaria Aldeneh, Vimal Thilak, Takuya Higuchi, Barry-John Theobald, Tatiana Likhomanenko}
\IEEEauthorblockA{\textit{Apple}\\
\{zaldeneh, vthilak\}@apple.com
}}

\begin{document}
%\ninept
%
\maketitle
\begin{abstract}
This study explores using embedding rank as an unsupervised evaluation metric for general-purpose speech encoders trained via self-supervised learning (SSL). Traditionally, assessing the performance of these encoders is resource-intensive and requires labeled data from the downstream tasks. Inspired by the vision domain, where embedding rank has shown promise for evaluating image encoders without tuning on labeled downstream data, this work examines its applicability in the speech domain, considering the temporal nature of the signals. The findings indicate rank correlates with downstream performance within encoder layers across various downstream tasks and for in- and out-of-domain scenarios. However, rank does not reliably predict the best-performing layer for specific downstream tasks, as lower-ranked layers can outperform higher-ranked ones. Despite this limitation, the results suggest that embedding rank can be a valuable tool for monitoring training progress in SSL speech models, offering a less resource-demanding alternative to traditional evaluation methods.

\end{abstract}
%
% \begin{keywords}
% \end{keywords}
%
\section{Introduction}
\label{sec:intro}
Self-supervised learning (SSL) enables learning from large amounts of unlabeled data to train general-purpose speech models~\cite{mohamed2022self,yang2024large}. Once trained, embeddings from these speech models can be used as features for downstream tasks, including automatic speech recognition, speaker identification, and keyword spotting. Evaluating the performance of SSL speech models on downstream tasks during their development, however, is a time-consuming and computationally expensive endeavor. These evaluations also presuppose the availability of labeled downstream data, which may not always be accessible. In addition, prior research has shown that the performance of the self-supervised training objective is not necessarily a reliable indicator of downstream task performance~\cite{chen2023reducing}. These challenges necessitate alternative evaluation methods for assessing the effectiveness of speech SSL models without requiring extensive downstream tuning and benchmarking.
% \textit{To this end, this work explores whether the rank of an SSL model's embedding indicates the model's quality on downstream tasks}.

Chen et al.~\cite{chen2023estimate} proposed two approaches to estimate the quality of pre-trained speech models for transfer learning for speech recognition tasks: one based on Bayesian likelihood estimation and the other related to optimal transport. These two approaches, however, relied on the availability of downstream data, while we seek measures that use \textit{only SSL pre-training data}. In the vision domain, several approaches~\cite{agrawal2022alphareq, garrido2023rankme, kalibhat2024measuring, lu2023using, thilak2024lidar, tsitsulin2023unsupervised} were proposed to estimate the quality of pre-trained models using only SSL pre-training data. Among these approaches, $\alpha$-Req~\cite{agrawal2022alphareq}, RankeMe~\cite{garrido2023rankme}, and LiDAR~\cite{thilak2024lidar} calculated metrics related to the spectrum of the embeddings to quantify the quality of representations. In~\cite{tsitsulin2023unsupervised}, the authors proposed the (in)coherence property from matrix sensing literature to quantify the quality of the embeddings. These methods estimated pre-trained model quality without relying on the downstream data or labels. However, the applicability of such methods to the speech domain remains unexplored in the literature.

In this work, we explore whether the rank of the embeddings from a speech SSL model provides an alternative measure of the model's quality, bypassing the need to run extensive downstream evaluations. In the vision domain, rank has been shown to be a simple and promising measure for estimating performance on downstream datasets without requiring labels or tuning for the downstream task~\cite{garrido2023rankme}. Rank measures the degree of dimensional collapse in a representation. Dimensional collapse is a phenomenon where the learned representations collapse into a lower-dimensional space that limits the expressive power of representations that in turn degrades performance measured on downstream tasks~\cite{hua2021feature}.
\textit{Despite rank's potential for assessing SSL models in the vision domain, whether it holds  for assessing   SSL models in the speech domain needs to be explored}. 

Evaluating speech SSL models on downstream tasks presents unique challenges that differentiate it from image classification. First, downstream speech tasks can be broadly divided into frame-level prediction tasks (i.e., tasks that require a sequence of labels for a given input sequence) and utterance-level prediction tasks (i.e., tasks that require a single label for a given input sequence). Second, downstream speech tasks focus on different components of speech, including but not limited to content (i.e., \textit{what is said?}) and speaker (i.e., \textit{who said it?}). Finally, prior work~\cite{yang2024large} has shown that different downstream tasks favor different layers from the speech SSL model (i.e., some tasks benefit from lower layers while others benefit from higher layers), while in image classification, the  penultimate layer is a conventional way to extract representations.

Our study considers the various factors encountered when evaluating SSL speech models, including task type (content or speaker), domain (in-domain or out-of-domain)  of pre-training data, and target (frame- or utterance-level predictions). 
We first extend rank of RankMe method~\cite{garrido2023rankme} to the temporal case by introducing RankMe-$t$.
Next, we assess the utility of rank both within individual layers and across multiple layers of the SSL speech model. Our results show that rank correlates with downstream performance within layers, and these correlations are consistent regardless of the pre-training data domain, target, and task type. Despite the within-layer correlations, our results show that rank cannot be used to pick the best layer for a given task (i.e., features from lower-rank layers can perform better on a downstream task than features from higher-rank layers). Regardless, our results suggest that rank can be employed for monitoring training, providing a less resource-intensive approach compared to traditional evaluation methods.

\section{Background: RankMe}
\label{sec:background}
If we have input samples $\{x_i\}_{i=0}^{n-1}$ and an encoder $f$ that gives a $d$-dimensional embedding $e_i\in\mathbb{R}^d$ given an input sample $x_i$, then we construct an embedding matrix $Z\in \mathbb{R}^{n\times d}$ by stacking the resulting embedding vectors. 
Then, RankMe~\cite{garrido2023rankme} defines the effective rank of $Z$, whose singular values are denoted as $\boldsymbol{\sigma}=[\sigma_1,\sigma_2,\ldots,\sigma_{min(n,d)}]$, as the entropy of the normalized singular values: 

\begin{equation}
    \text{RankMe}\left(Z\right)=\exp\left( -\sum_i p_i\log p_i \right),
    \label{eqn:rankme}
\end{equation}
where
\begin{equation}
    p_i=\frac{\sigma_i}{\|\boldsymbol{\sigma}\|_1}.
    \label{eqn:rankme:normalization}
\end{equation}

% $$\text{RankMe}\left(Z\right)=\exp\left( -\sum_i p_i\log p_i \right),\quad p_i=\frac{\sigma_i}{\|\sigma\|_1}.$$

\section{RankMe-\texorpdfstring{$t$}:: RankMe Temporal Extension}
\label{sec:approach}
In the case of speech SSL models, the input samples are sequences $x_i=[x^1_i,x^2_i,\ldots,x^{T_i}_i]$, where $T_i$ is the length of the $i$-th sample in the dataset. The encoder at the $l$-th layer produces a sequence of embedding vectors $e_i=[e^{1}_i,e^{2}_i,\ldots,e^{T'_i}_i]$, with $T'_i \leq T_i$, given an input sequence. We apply zero-padding to all embedding sequences to match the length $T'_{max}$, the longest embedding sequence in the dataset. The resulting padded embedding sequences are then stacked to form the final matrix sequence $Z=[Z^1, Z^2,\ldots, Z^{T'_{max}}]$. We extend RankMe~\cite{garrido2023rankme} to the temporal case  and compute the rank over the sum of the embedding matrices:
\begin{equation}
    \text{RankMe-}t(Z) = \text{RankMe}(Z^1 + Z^2 + \ldots + Z^{T'_{max}}).
    \label{eqn:rankme-t}
\end{equation}
% $$\text{RankMe-}t(Z) = \text{RankMe}(Z^1 + Z^2 + \ldots + Z^{T'_i}).$$
% Note that this extension exploits the basic additive property of the rank, i.e., for any matrices $A, B\in\mathbb{R}^{n\times d}$, $rank(A + B) \leq rank(A) + rank(B)$. 
Note that the sum of the embedding matrices in Equation~(\ref{eqn:rankme-t}) depends on the sequence length $T'_{max}$ and one should consider normalized sum by dividing the sum to the sequence length~$T'_{max}$. However, the resulting rank in Equation~(\ref{eqn:rankme-t}) is independent from the sequence length $T'_{max}$ as RankMe uses normalized singular values, see Equation~(\ref{eqn:rankme:normalization}).
Also note, that our definition is consistent with utterance-level representations: often for utterance-level downstream tasks (e.g., speaker identification) mean pooling is applied to induce a fixed-size embedding. The latter is the same as RankMe for image classification.
Alternatively, one could compute $\text{RankMe-}t$ by first summing the sequence of embedding vectors for each sample, $\sum_{t=1}^{T'_{i}} e^{t}_i$, to obtain a fixed-size embedding vector. Once this is done for each sample, the resulting embedding vectors can be stacked to form the matrix $Z$.

\textbf{\textit{Our goal is to study the relationship between the $\text{RankMe-}t$ values and the corresponding downstream performances for different tasks and different model configurations.}
}
% We take the mean across time before computing the rank of the resulting matrix and study the relationship between this rank value and the downstream performance for different tasks and model configurations.

The input samples used to calculate \emph{RankMe-$t$} are from pre-training data used to optimize the SSL model, and thus, the proposed measure does not rely on a separate validation set or labels. The use of a source dataset is justified via arguments in~\cite{garrido2023rankme} that show that the training accuracy on a classification or regression tasks improves with embedding rank. Furthermore, Garrido et al.~\cite{garrido2023rankme} conducted empirical analysis with image-based SSL models to confirm that large effective rank is highly correlated with downstream task performance. Simon et al.~\cite{simon2023stepwise} studied SSL methods theoretically in a simplified setting and showed that the rank of the representations increases in a step-wise manner during optimization. This result suggests that the hyper-parameters for optimizing SSL methods should be chosen to maximize the embedding rank at the end of training. Taking the above observations into consideration, we hypothesize that the embedding rank is a useful measure of representation quality in speech SSL methods too. Therefore, we propose to use \emph{RankMe-$t$} described in Equation~$\left(\ref{eqn:rankme-t}\right)$ to quantify speech representation quality in this paper.

\begin{table}
  \centering
  \caption{A description of the four downstream tasks that we study. PR denotes the phoneme recognition task on the LibriSpeech dataset; TIMIT denotes the phoneme recognition task on the TIMIT dataset; KS denotes the keyword spotting task on the Speech Commands dataset; SID denotes the speaker identification task on the VoxCeleb$1$ dataset. ``Target'' indicates if the task is sequence- or utterance-based. ``Type'' indicates if the task focuses on the content or the speaker.}
  \label{table_tasks}
  \begin{tabular}{lccc}
    \toprule
    \textbf{Task} & \textbf{Domain} & \textbf{Target} & \textbf{Type} \\
    \midrule
    PR & In-domain & Sequence & Content \\
    TIMIT & Out-of-domain & Sequence & Content \\
    KS & Out-of-domain & Utterance & Content \\
    SID & Out-of-domain & Utterance & Speaker \\   
    \bottomrule
  \end{tabular}
\end{table}

\section{Setup}
\label{sec:setup}
\subsection{SSL Model Training}
\label{sec:setup:ssl_model_training}
We use the HuBERT Base model~\cite{hsu2021hubert} in our study because it is widely used (due to open-sourced models) and was used for training state-of-the-art WavLM model~\cite{chen2022wavlm}\footnote{We use the official \texttt{PyTorch} implementation: \url{https://github.com/pytorch/audio/tree/main/examples/hubert}.}. We use the LibriSpeech $960$ hours dataset~\cite{panayotov2015librispeech} to train nine configurations of the first iteration of HuBERT on $32$ GPUs using the following hyper-parameters: number of MFCC clusters: $\{50,100,500\}$; masking probability: $\{6.5\%, 8\%, 9\%\}$. We train all models for $300$k steps and use the default learning rate of $5e\text{-}4$ and batch size of $87.5$ seconds per GPU. We save a checkpoint every $20$k steps and use these checkpoints for downstream evaluation. The HuBERT Base model consists of $12$ transformer layers, with each layer outputting a sequence of $768$-dimensional embeddings.

\subsection{Downstream Model Training}
% \newpara{Downstream Model Training.}
Given a trained HuBERT checkpoint, we evaluate the embeddings extracted from 
layers $\{0, 3, 6, 9\}$ on four downstream tasks: phoneme recognition on LibriSpeech~\cite{panayotov2015librispeech}; phoneme recognition on TIMIT~\cite{garofolo1993darpa}; keyword spotting on Speech Commands~\cite{warden2017speech}; and speaker identification on the VoxCeleb1~\cite{nagrani2017voxceleb}. We limit the evaluation to these four layers to reduce computational burden while maintaining sufficient coverage across the network's depth.
A more detailed description of the tasks is provided in Table~\ref{table_tasks}. With the exception of TIMIT, where we use $50$k training steps when training the downstream model, we use the default hyper-parameter setups from the \texttt{S3PRL} toolkit\footnote{\texttt{S3PRL} toolkit: \url{https://github.com/s3prl/s3prl}.} for all downstream models. Performance on the LibriSpeech phoneme recognition is measured using phoneme error rate (PER);  the performance on the remaining tasks is measured using accuracy.

\subsection{Rank Measurement}
% \newpara{Rank Measurement.}
Given a trained HuBERT checkpoint, we compute RankMe-$t$ of the sequence of representations from a specific layer that results from passing $10$k randomly sampled utterances from the LibriSpeech \texttt{train-clean} set. Note that RankMe-$t$ is computed using the same $10$k randomly sampled utterances of LibriSpeech regardless of the downstream task being assessed. The goal is to study the relationship between the resulting rank and the downstream performance achieved by a checkpoint and layer. We use Kendall’s $\tau$ rank correlation coefficient to measure the relationship between the pairs of measurements.

\section{Results}
\begin{figure}[t]
    \centering
\hspace{15pt}\def\svgwidth{0.9\columnwidth}\footnotesize%% Creator: Inkscape 1.3.2 (091e20e, 2023-11-25), www.inkscape.org
%% PDF/EPS/PS + LaTeX output extension by Johan Engelen, 2010
%% Accompanies image file 'figures2/summary-all-legend.3.pdf' (pdf, eps, ps)
%%
%% To include the image in your LaTeX document, write
%%   \input{<filename>.pdf_tex}
%%  instead of
%%   \includegraphics{<filename>.pdf}
%% To scale the image, write
%%   \def\svgwidth{<desired width>}
%%   \input{<filename>.pdf_tex}
%%  instead of
%%   \includegraphics[width=<desired width>]{<filename>.pdf}
%%
%% Images with a different path to the parent latex file can
%% be accessed with the `import' package (which may need to be
%% installed) using
%%   \usepackage{import}
%% in the preamble, and then including the image with
%%   \import{<path to file>}{<filename>.pdf_tex}
%% Alternatively, one can specify
%%   \graphicspath{{<path to file>/}}
%% 
%% For more information, please see info/svg-inkscape on CTAN:
%%   http://tug.ctan.org/tex-archive/info/svg-inkscape
%%
\begingroup%
  \makeatletter%
  \providecommand\color[2][]{%
    \errmessage{(Inkscape) Color is used for the text in Inkscape, but the package 'color.sty' is not loaded}%
    \renewcommand\color[2][]{}%
  }%
  \providecommand\transparent[1]{%
    \errmessage{(Inkscape) Transparency is used (non-zero) for the text in Inkscape, but the package 'transparent.sty' is not loaded}%
    \renewcommand\transparent[1]{}%
  }%
  \providecommand\rotatebox[2]{#2}%
  \newcommand*\fsize{\dimexpr\f@size pt\relax}%
  \newcommand*\lineheight[1]{\fontsize{\fsize}{#1\fsize}\selectfont}%
  \ifx\svgwidth\undefined%
    \setlength{\unitlength}{384.57000732bp}%
    \ifx\svgscale\undefined%
      \relax%
    \else%
      \setlength{\unitlength}{\unitlength * \real{\svgscale}}%
    \fi%
  \else%
    \setlength{\unitlength}{\svgwidth}%
  \fi%
  \global\let\svgwidth\undefined%
  \global\let\svgscale\undefined%
  \makeatother%
  \begin{picture}(1,0.07364068)%
    \lineheight{1}%
    \setlength\tabcolsep{0pt}%
    \put(0,0){\includegraphics[width=\unitlength,page=1]{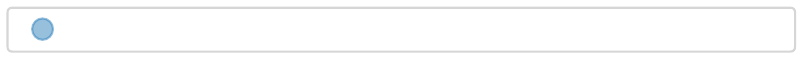}}%
    \put(0.10921289,0.02995553){\makebox(0,0)[lt]{\lineheight{1.25}\smash{\begin{tabular}[t]{l}Layer 0\end{tabular}}}}%
    \put(0,0){\includegraphics[width=\unitlength,page=2]{figures2/summary-all-legend.3.pdf}}%
    \put(0.35516128,0.02995553){\makebox(0,0)[lt]{\lineheight{1.25}\smash{\begin{tabular}[t]{l}Layer 3\end{tabular}}}}%
    \put(0,0){\includegraphics[width=\unitlength,page=3]{figures2/summary-all-legend.3.pdf}}%
    \put(0.60110967,0.02995553){\makebox(0,0)[lt]{\lineheight{1.25}\smash{\begin{tabular}[t]{l}Layer 6\end{tabular}}}}%
    \put(0,0){\includegraphics[width=\unitlength,page=4]{figures2/summary-all-legend.3.pdf}}%
    \put(0.84705806,0.02995553){\makebox(0,0)[lt]{\lineheight{1.25}\smash{\begin{tabular}[t]{l}Layer 9\end{tabular}}}}%
  \end{picture}%
\endgroup%

    \vspace{-10pt}
    \def\svgwidth{0.9\columnwidth}
    \footnotesize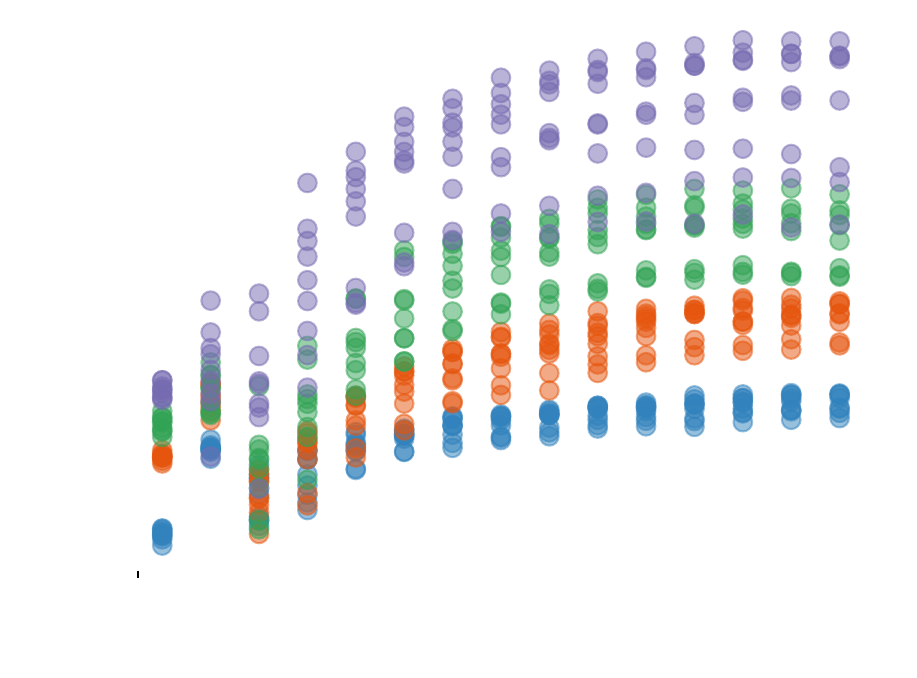
    \vspace{0.2cm}
    \caption{The rank ($y$-axis), estimated via RankMe-$t$, dependence on the training step ($x$-axis) for a given layer when training HuBERT. Different points for a given layer and step represent different hyper-parameter settings described in Section~\ref{sec:setup:ssl_model_training}.}
    \label{fig:fig_1}
\end{figure}

Figure~\ref{fig:fig_1} shows how the RankMe-$t$ of constructed embeddings from HuBERT changes with training steps. We find that the \textbf{RankMe-$t$ increases with the training step}, and the rate of increase is higher earlier in the training process than later in the process. Deeper layers start with higher average effective ranks compared to earlier layers; layers zero, three, six, and nine give an average effective rank of $158.25$, $229.20$, $260.61$, and $289.77$, respectively, after training for around $10$k steps. The RankMe-$t$ of a given training step and layer can vary with the hyper-parameter choice. For instance, the representations from layer nine at the last training step give an effective rank of $441.20$ when training HuBERT with $500$ clusters and $9\%$ masking probability, but give a rank of $608.37$ when training with $50$ clusters and $8\%$ masking probability. However, this \textbf{range of the ranks from hyper-parameter setups is inconsistent across layers}---lower layers show a smaller range compared to higher layers. Also, we observe that the order of the setups is different per layer, i.e., the setup that gives the highest rank for layer nine is not necessarily the same setup that gives the highest rank for layer three.

\begin{figure*}[htbp]
    \centering
\hspace{15pt}\def\svgwidth{0.9\columnwidth}
        \footnotesize
    \vspace{-10pt}
    \subfloat[Phoneme Recognition (PR)]{
        \def\svgwidth{0.9\columnwidth}
        \footnotesize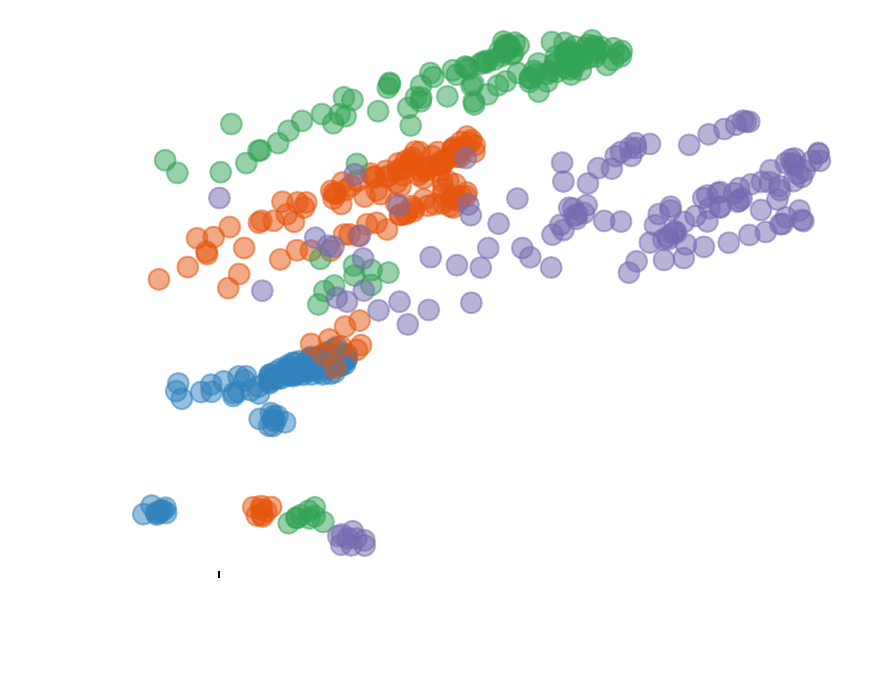
        \label{fig:subfig1}
    }
    % \hfill
    \subfloat[Phoneme Recognition (TIMIT)]{
        \def\svgwidth{0.9\columnwidth}
        \footnotesize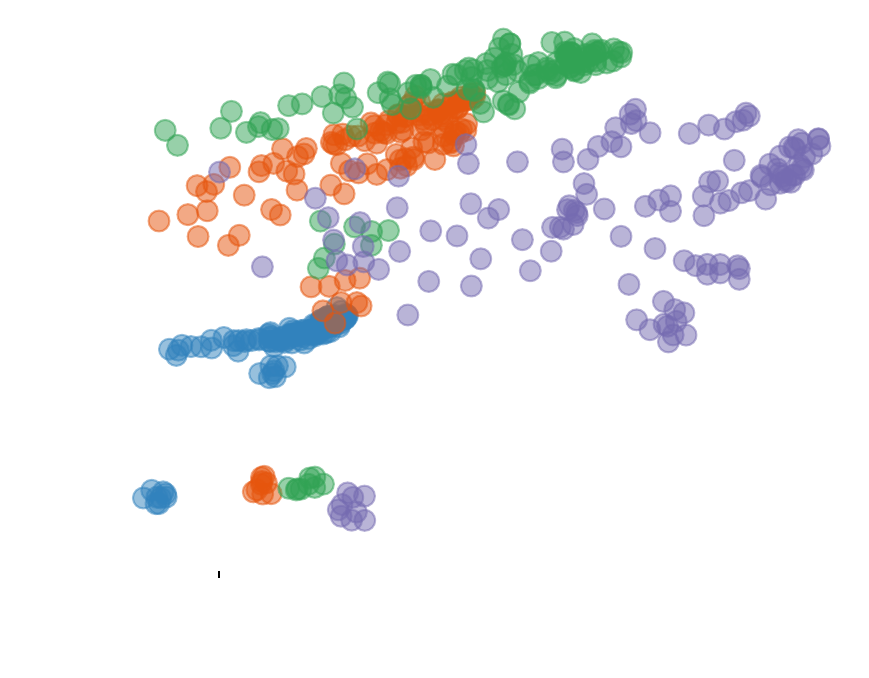
        \label{fig:subfig2}
    }\\
    \vspace{15pt}
    % \hfill
    \subfloat[Keyword Spotting (KS)]{
        \def\svgwidth{0.9\columnwidth}
        \footnotesize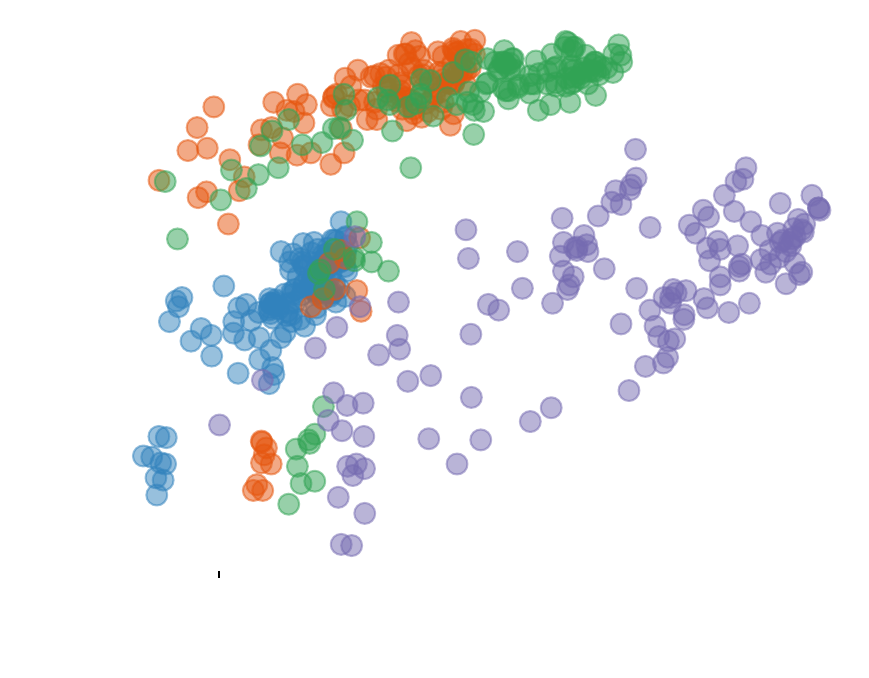
        \label{fig:subfig3}
    }
    % \hfill
    \subfloat[Speaker Identification (SID)]{
        \def\svgwidth{0.9\columnwidth}
        \footnotesize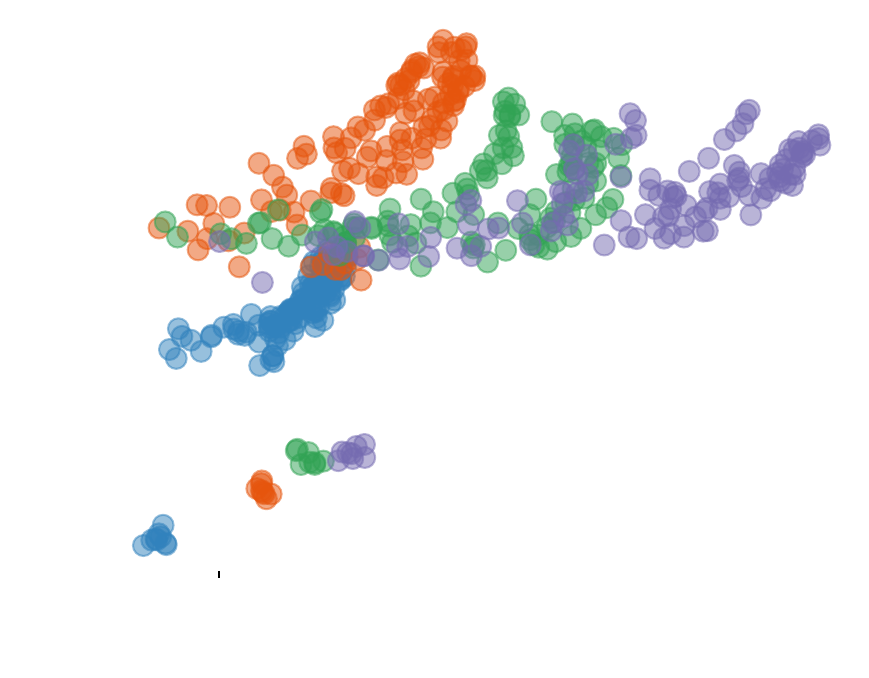
        \label{fig:subfig4}
    }
    \caption{The figure shows a positive correlation between the downstream performance (y-axis) and the embedding rank (x-axis) estimated via RankMe-$t$ for each layer ($0$th, $3$rd, $6$th, $9$th) on the four tasks described in Table~\ref{table_tasks}. Different points for the same layer correspond to different hyper-parameters settings for SSL training. 
    Despite the positive correlations that we observe per layer, we cannot use rank to predict which layer performs the best on a downstream task.
    }
    \label{fig:fig_2}
\end{figure*}

Figure~\ref{fig:fig_2} shows how the rank of constructed embeddings from HuBERT relates to the performance on the four downstream tasks detailed in Table~\ref{table_tasks}. Overall, we find \textbf{significant ($p \ll 1e\text{-}10$) positive correlations between RankMe-$t$ and downstream performances for all four tasks} (PR: $\tau=0.49$; TIMIT: $\tau=0.44$; KS: $\tau=0.29$; SID: $\tau=0.45$), suggesting that the correlation persists regardless of the pre-training data domain (LibriSpeech vs. TIMIT), the target (sequence vs. utterance), and the downstream task type (content vs. speaker). Table~\ref{table_layer_correlations} shows that these significant positive correlations exist when looking at representations from each layer independently. \textbf{\textit{In other words, when the rank of an embedding from a layer increases, the performance of that layer on downstream task also tends to improve.}}

The results in Figure~\ref{fig:fig_2} show that different downstream tasks prefer features from different layers; for instance, layer six gives the best performance on phoneme recognition, layers three and six give the best performance on keyword spotting, and layer three gives the best performance on the speaker identification. These observations are consistent with observations in prior works~\cite{chen2022wavlm,yang2024large}. We observe that \textbf{the effective rank of an embedding is not indicative of layer performance on a specific task}. For example, as it is observed in Figure~\ref{fig:subfig4}, features from layers six and nine have higher ranks despite yielding lower accuracy on speaker identification compared to features from layer three.

\begin{table}[t]
  \centering
  \caption{Kendall’s $\tau$ coefficient between ranks, estimated via RankMe-$t$, and downstream performance within layers. A description of the four tasks is provided in Table~\ref{table_tasks}. All shown correlations are statistically significant ($p \ll 1e\text{-}10$).}
  \label{table_layer_correlations}
  \begin{tabular}{lcccc}
    \toprule
    \textbf{Layer} & \textbf{PR} & \textbf{TIMIT} & \textbf{KS} & \textbf{SID}\\
    \midrule
    $0$ & $0.67$ & $0.75$ & $0.57$  & $0.73$\\
    $3$ & $0.56$ & $0.60$ & $0.56$  & $0.64$\\
    $6$ & $0.61$ &  $0.64$ & $0.58$ & $0.41$ \\
    $9$ & $0.43$  & $0.34$ & $0.46$ & $0.56$ \\
    \bottomrule
  \end{tabular}
\end{table}

\section{Discussion \& Conclusion }
\label{sec:discussion}
We sought to study rank's potential as a proxy metric for assessing the quality of features from SSL speech models. 
To this end, we first proposed RankMe-$t$, a simple extension of RankMe~\cite{garrido2023rankme} to the temporal case encountered in the speech domain. We then trained  HuBERT models with nine different configurations and saved checkpoints along the way. We computed the rank of features from a given layer for each checkpoint and evaluated the features' performance on four downstream tasks. Our results showed that rank correlates with downstream performance within layers, and these correlations were consistent regardless of the pre-training data domain (LibriSpeech vs. TIMIT), target (sequence vs. utterance), and downstream task type (content vs. speaker). However, our results showed that \textbf{rank cannot be used to pick the best layer for a given task} (i.e., features from lower-rank layers can perform better on a downstream task compared to features from higher-rank layers). 
We believe, something more fundamental is hidden in how embeddings are learnt between the layers in speech SSL models, particularly in HuBERT models, and
we hope, our findings will further stimulate research in that direction to uncover missing pieces in addition to the rank.
\textit{Regardless, our results suggest that rank can be employed for monitoring SSL training and selecting the checkpoints which provide the improved ranks of the embeddings across all layers.}

% \vfill\pagebreak

\bibliographystyle{IEEEtran}
\bibliography{IEEEabrv,refs}

\end{document}